\documentclass[conference]{IEEEtran}
\IEEEoverridecommandlockouts
\usepackage{cite}
\usepackage{amsmath,amssymb,amsfonts}
\usepackage{algorithm}
\usepackage{algorithmic}
\usepackage{graphicx}
\usepackage{textcomp}
\usepackage{xcolor}
\usepackage{booktabs}
\usepackage{tabularx}
\usepackage{url}
\def\BibTeX{{\rm B\kern-.05em{\sc i\kern-.025em b}\kern-.08em
    T\kern-.1667em\lower.7ex\hbox{E}\kern-.125emX}}

\begin{document}

\title{LoopVSR: A Loop Engineering Framework for Automated Repair of Visual Speech Recognition Inference Pipelines}

\author{
\IEEEauthorblockN{
Fei Qin\IEEEauthorrefmark{1},
Bowen Zhang\IEEEauthorrefmark{2}\IEEEauthorrefmark{3},
Chao Fan\IEEEauthorrefmark{2}\IEEEauthorrefmark{3},
Pengcheng Luo\IEEEauthorrefmark{2}\IEEEauthorrefmark{3}, and
Genke Yang\IEEEauthorrefmark{2}\IEEEauthorrefmark{3}}
\IEEEauthorblockA{\IEEEauthorrefmark{1}
\textit{Department of Traffic and Public Security Detention Facilities Administration},
\textit{Shanghai Police College}\\
Shanghai, China\\
2979729486@qq.com}
\IEEEauthorblockA{\IEEEauthorrefmark{2}
\textit{Ningbo Artificial Intelligence Institute},
\textit{Shanghai Jiao Tong University}, Ningbo, China}
\IEEEauthorblockA{\IEEEauthorrefmark{3}
\textit{School of Automation and Intelligent Sensing},
\textit{Shanghai Jiao Tong University}, Shanghai, China}
\IEEEauthorblockA{\{bwz96sco,fchao2025,luopeng69131,gkyang\}@sjtu.edu.cn}
\thanks{The source code is available at \protect\url{https://github.com/luopeng69131/LoopVSR}.}
}

\maketitle

\begin{abstract}
Visual speech recognition (VSR) recovers speech from lip movements when audio is noisy or unavailable. Its multi-stage inference pipeline spans video decoding, mouth-region extraction, preprocessing, model invocation, and decoding, where upstream failures can mask downstream faults. Pipeline maintenance therefore still relies largely on predefined checks and manual debugging. We propose LoopVSR, a Loop Engineering framework that enables a code agent to automatically diagnose and repair VSR inference pipelines using end-to-end execution evidence. It couples constrained repository-level diagnosis and patching with an external controller that audits changes, runs real inference, and accepts or rolls back patches using failures and character error rate (CER). The resulting feedback loop returns newly observed exceptions, tensor statistics, and recognition errors to the agent, progressively exposing faults masked by upstream failures. On the CMLR VSR system, LoopVSR repairs all 11 main faults with 100\% mean recovery, whereas the Static guard repairs 2 of 11 with 18.13\% mean recovery. It also resolves three cascading tasks in seven accepted iterations and preserves recovery on an independent 200-video hidden set. These results demonstrate that LoopVSR enables measurable, end-to-end automated repair of VSR inference pipelines.
\end{abstract}


\section{Introduction}

Visual speech recognition (VSR) recovers spoken content from lip movements and provides a complementary channel when audio is noisy or unavailable. Advances in audio-visual data, visual encoders, and sequence modeling have substantially improved recognition accuracy, moving VSR from model feasibility toward deployable systems~\cite{shi2022avhubert,ma2022multilingualvsr}.

A deployed VSR system relies on a multi-stage inference pipeline that typically contains video decoding, frame-rate handling, facial landmark detection, mouth region-of-interest (ROI) extraction, normalization, temporal transforms, model invocation, and search decoding. Errors in any of these stages can degrade recognition quality. Such faults may span source code and configuration; their symptoms need not identify their root causes, and an upstream runtime failure can hide a downstream quality error.

\begin{figure}[t]
\centering
\includegraphics[width=\columnwidth]{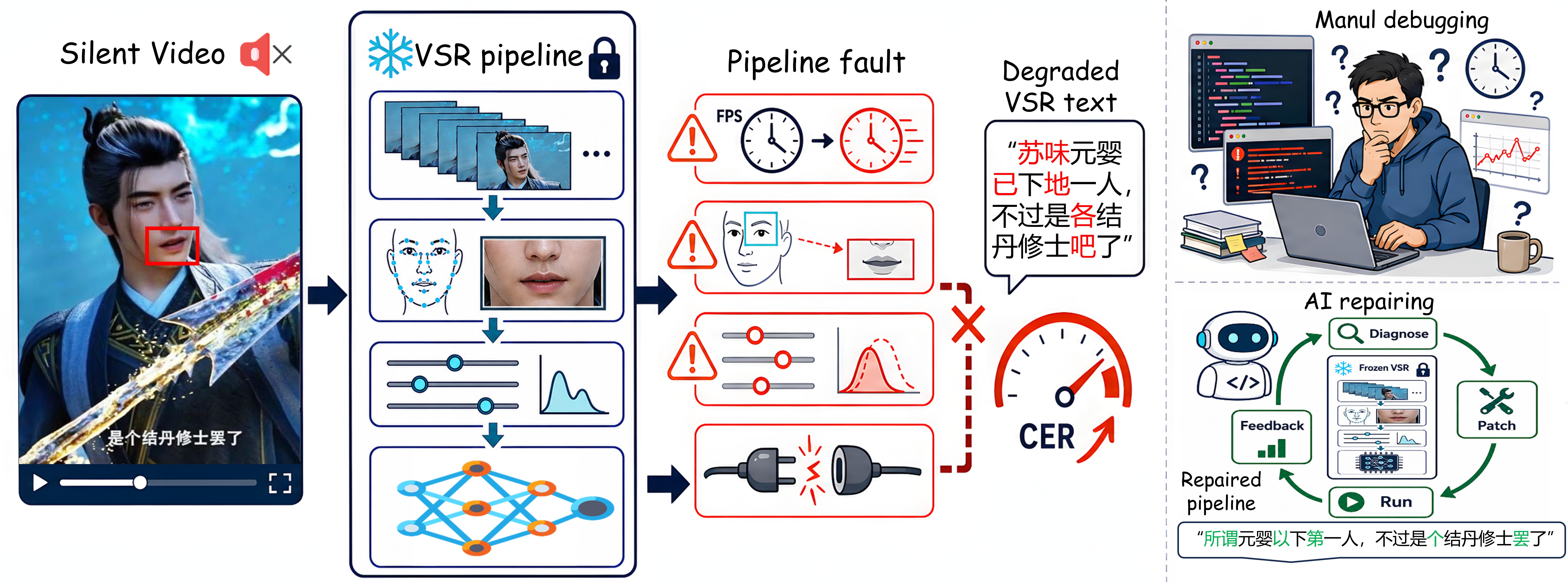}
\caption{Motivation for LoopVSR. Manual maintenance requires engineers to trace failures across multiple VSR stages. Loop Engineering instead lets engineers define the repair objective, runtime evidence, evaluation, and control rules that guide a coding agent toward a verified pipeline repair.}
\label{fig:motivation}
\end{figure}

Existing VSR studies primarily improve representation learning, model architectures, and training data~\cite{shi2022avhubert,ma2022multilingualvsr,ma2023autoavsr}, leaving inference-pipeline repair outside their scope. Static rules can detect predefined anomalies, but cannot readily infer cross-stage causes from end-to-end recognition degradation. Although automated program repair can evaluate patches through executable tests~\cite{legoues2012genprog,nguyen2013semfix}, a local or one-shot correction does not establish that VSR recognition quality has recovered. What is missing is an automated process that connects diagnosis and code modification to repeated real inference and measurable task feedback.

Large language model coding agents, such as OpenAI Codex~\cite{openai2025codex} and Anthropic Claude Code~\cite{anthropic2025claudecode}, can reason over repositories, modify code, invoke tools, and act on execution results~\cite{yang2024sweagent,bouzenia2025repairagent}. These capabilities motivate a shift from turn-by-turn prompting to \emph{Loop Engineering}: the engineer designs the goal, execution context, tools, verification, persistent state, operational boundaries, and stopping conditions of an agentic system, enabling it to repeatedly prompt, check, and redirect the agent with limited human intervention~\cite{anthropic2026loops}. In VSR, this perspective recasts pipeline maintenance from manually tracing individual failures to engineering the evidence and control mechanisms through which an agent can pursue and verify an end-to-end repair. Fig.~\ref{fig:motivation} illustrates this motivation.

Building on this motivation, we propose \textbf{LoopVSR}, a Loop Engineering framework for automated repair of VSR inference pipelines. LoopVSR instantiates the engineering loop with an isolated editable workspace, protected execution and evaluation components, runtime observations, patch auditing, accept-or-rollback control, and an explicit stopping condition. The code agent modifies source code and configuration using exceptions, tensor statistics, recognition cases, and character error rate (CER) feedback. Re-execution can expose previously masked faults, allowing the agent to repair compound failures over multiple iterations.

The main contributions of this work are summarized as follows:
\begin{itemize}
    \item We formulate automated repair of VSR inference pipelines and introduce a Loop Engineering workflow driven by real execution feedback.
    \item We design an auditable protocol covering configuration, preprocessing, ROI, temporal, runtime-interface, and compound faults while protecting model weights, evaluation logic, and hidden data.
    \item On a CMLR system, LoopVSR repairs all 11 faults with 100\% mean recovery, while a static guard repairs 2 of 11; it also resolves three cascading faults and preserves performance on hidden evaluation.
\end{itemize}

\section{Related Work}

VSR research has progressed from end-to-end sentence-level lipreading with LipNet~\cite{assael2016lipnet} to self-supervised audio-visual representation learning with AV-HuBERT~\cite{shi2022avhubert}. Multilingual VSR improves model design, augmentation, and cross-lingual data use~\cite{ma2022multilingualvsr}, while Auto-AVSR enlarges training sets through automatic transcriptions~\cite{ma2023autoavsr}. These studies substantially improve recognition models and training data, but do not address automatic repair of the surrounding inference pipeline, including video processing, mouth ROI construction, runtime interfaces, and decoding.

Addressing these pipeline failures requires software-level diagnosis and modification. Automated program repair uses executable feedback to search or synthesize patches, as represented by GenProg and SemFix~\cite{legoues2012genprog,nguyen2013semfix}; DeepDiagnosis extends fault diagnosis to numerical states in deep-learning training~\cite{wardat2022deepdiagnosis}. More recently, ReAct and Reflexion enable language agents to interleave reasoning, action, and feedback~\cite{yao2023react,shinn2023reflexion}, while SWE-bench, SWE-agent, AutoCodeRover, and RepairAgent apply these capabilities to repository-level software tasks~\cite{jimenez2024swebench,yang2024sweagent,zhang2024autocoderover,bouzenia2025repairagent}. However, these methods are primarily guided by software tests or training-time signals rather than task-level behavior of a VSR system. LoopVSR instead audits agent patches through external video inference and combines failure counts, tensor telemetry, and CER with accept-or-rollback decisions, forming a domain-specific and measurable repair loop.

\section{Problem Formulation}

We study engineering-pipeline repair for a deployed VSR system. Given an input video $v$, a complete VSR system is written as
\begin{equation}
    \hat{y}=F(v;W,P,C),
\end{equation}
where $\hat{y}$ is the predicted text, $W$ denotes frozen model weights, $P$ denotes pipeline source code for video processing, feature construction, and inference, and $C$ denotes runtime configurations such as frame rate, model path, and decoding parameters. Conventional model optimization updates $W$ to improve recognition. In contrast, we keep $W$ fixed and permit changes only to fault-related portions of $P$ and $C$. This setting reflects deployment scenarios in which the model has already been trained and validated, but its surrounding processing or runtime configuration has become corrupted.

Let the system state at repair iteration $t$ be
\begin{equation}
    x_t=(P_t,C_t).
\end{equation}
An external evaluator executes the current system on a repair-development set $\mathcal{D}_r$ and returns
\begin{equation}
    o_t=E(x_t,\mathcal{D}_r)
       =\{f_t,\mathrm{CER}_t,z_t,e_t,q_t\},
\end{equation}
where $f_t$ is the number of failed samples, $\mathrm{CER}_t$ is the corpus-level CER, $z_t$ contains input tensor shapes and numerical statistics, $e_t$ contains initialization or runtime exceptions, and $q_t$ contains representative high-error samples. We compute CER as
\begin{equation}
    \mathrm{CER}_t=
    \frac{\sum_i(S_i+I_i+D_i)}
         {\sum_i N_i},
\end{equation}
where $S_i$, $I_i$, and $D_i$ are the numbers of character substitutions, insertions, and deletions for sample $i$, respectively, and $N_i$ is its reference length.

The code agent generates a patch from the current state, external observation, and repair history:
\begin{equation}
    \Delta_t=A(x_t,o_t,\mathcal{H}_t),
\end{equation}
yielding a candidate state $x'_t=x_t\oplus\Delta_t$. The goal is not to reproduce a predetermined source-code snapshot. With frozen model weights and constrained edit boundaries, the objective is to eliminate runtime failures and recover CER to the healthy system's level. LoopVSR therefore judges repairs by real end-to-end inference rather than textual patch similarity.

\section{Proposed LoopVSR Framework}

LoopVSR realizes Loop Engineering as a bounded repair system around a code agent. It separates repository-level diagnosis and modification from evaluation authority: the agent proposes changes, while an external controller audits each patch, measures its effect through real inference, and maintains the accepted pipeline state. Repeated execution supplies the evidence needed to retain, roll back, or refine a repair without relying on the agent's own completion judgment.

\subsection{Overall Architecture}

\begin{figure}[t]
\centering
\includegraphics[width=\columnwidth]{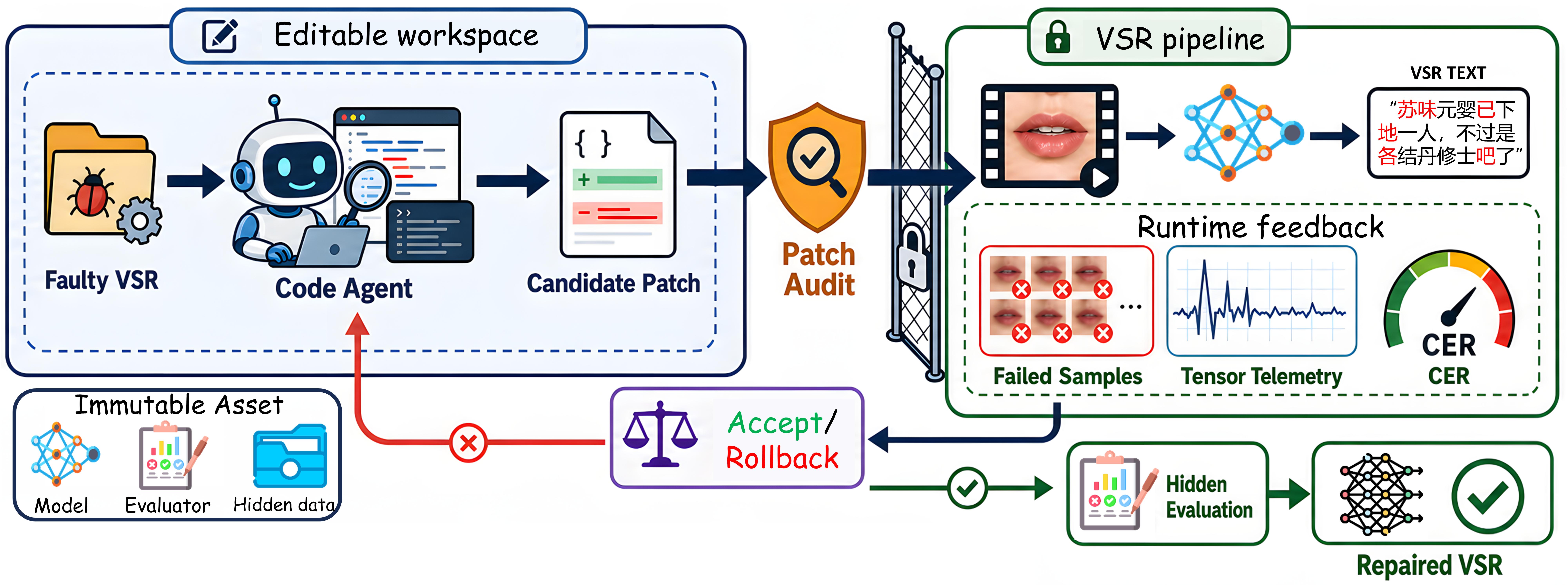}
\caption{Architecture of LoopVSR. A code agent repairs the faulty implementation inside an editable workspace. Audited patches are executed by the protected VSR pipeline, and runtime feedback controls acceptance or rollback; model weights, the evaluator, and hidden data remain immutable.}
\label{fig:architecture}
\end{figure}

LoopVSR connects an editable workspace and code agent to protected patch auditing, VSR execution, and evaluation. The agent produces a candidate patch from the faulty implementation and accumulated feedback; the controller then audits and executes it, returning failed-sample counts, tensor telemetry, and CER. These measurements govern acceptance and rollback and become context for the next iteration. Model weights, the evaluator, and hidden data remain immutable throughout the loop, preventing the repair process from changing its own target or evidence. The resulting architecture and information flow are shown in Fig.~\ref{fig:architecture}.

\subsubsection{Editable Workspace}

Each task starts from a designated repository revision in an isolated Git workspace containing the editable inference source and configuration, task constraints, current feedback, and repair history. The faulty initial state is committed before agent execution, and every accepted patch produces a new checkpoint. An ineffective candidate can therefore be discarded without contaminating later iterations. The editable boundary covers video reading, landmark processing, ROI construction, spatial and temporal transformations, model invocation, and decoding. The healthy source, model files, evaluator, data manifests, and hidden set remain outside the workspace. This organization provides both an exact repair trajectory and a reproducible rollback point.

\subsubsection{Code Agent}

At each iteration, the code agent reads the current implementation, task constraints, repair history, and controller feedback, then formulates one primary fault hypothesis and implements a minimal testable patch. Restricting an iteration to one principal hypothesis makes the measured effect attributable to a specific diagnosis while still permitting coordinated edits across related files. The agent may search code, inspect stack traces, compare configuration values, and reason over tensor shapes or distributions with local tools, but has no network, external file, remote model, or evaluator access. For example, a shortened input sequence directs attention to frame-rate or temporal-sampling logic, whereas a shifted numerical range motivates inspection of scaling and normalization. The hypothesis and resulting patch are retained in the history for subsequent reasoning.

\subsubsection{Protected VSR Execution and Loop Control}

The controller synchronizes only audited source and configuration changes to an independent graphics processing unit (GPU) environment and runs the complete VSR pipeline with fixed model weights and evaluation logic. Because the executor, metric implementation, and data manifest are outside the agent workspace, reported improvement must arise from actual pipeline behavior. The controller maintains candidate and accepted states, execution artifacts, evaluation results, and decision records. If the repair target is not met, it converts the newly observed evidence into structured feedback and starts another agent iteration; after termination, it finalizes the accepted state for hidden evaluation.

\subsection{Runtime Feedback}

Runtime feedback connects local pipeline states to end-to-end recognition quality. For each run, the executor aggregates initialization and sample-level exceptions, failed-sample counts, edit errors, corpus CER, and representative high-error predictions. It also reports the pre-model tensor dimensions, data type, minimum, maximum, mean, and standard deviation for selected inputs. These signals are complementary: exceptions identify interface or dependency failures; temporal and spatial dimensions expose frame-rate, ROI, and sampling faults; numerical ranges reveal scaling or normalization shifts; and recognition cases and CER determine whether execution recovery also restores task quality.

Feedback follows runtime observability rather than a predetermined fault label. The controller returns only evidence produced by stages reached in the current execution. A loader failure, for example, initially hides tensors and predictions; once loading is repaired, the next run can expose an incorrect tensor distribution or downstream CER degradation. The loop therefore expands its diagnostic evidence as the pipeline recovers, while healthy source code, correct patches, and hidden examples remain undisclosed.

\subsection{Patch Audit and Safety Constraints}

Before execution, LoopVSR requires a non-empty patch confined to permitted inference source or configuration files. The auditor inspects the complete diff and rejects model-path or parameter changes, modifications to task specifications or evaluation assets, and patches containing sample identifiers, reference text, or stored predictions. The restricted workspace also blocks network and out-of-workspace access. Only compliant candidates reach the executor; all others are recorded with a rejection reason and rolled back. These constraints prevent metric modification, transcript hard-coding, and weight replacement, separating the agent's ability to generate code from the controller's authority to validate a repair.

\subsection{Accept-or-Rollback Control and Termination}

The controller prioritizes runtime validity: it retains an audited candidate when the failure count decreases, or when the failure count is unchanged and CER improves by more than $\epsilon$; otherwise it restores the latest accepted state. The margin $\epsilon$ suppresses changes caused by numerical noise. Failure reduction is accepted even before CER improves because an upstream interface repair can make previously unreachable stages observable; subsequent iterations must then recover recognition quality. Thus, every accepted transition improves the lexicographically ordered pair of failure count and CER.

Performance recovery is
\begin{equation}
R_t=
\frac{\mathrm{CER}_{\mathrm{fault}}-\mathrm{CER}_t}
     {\mathrm{CER}_{\mathrm{fault}}-\mathrm{CER}_{\mathrm{clean}}},
\end{equation}
where $\mathrm{CER}_{\mathrm{fault}}$ and $\mathrm{CER}_{\mathrm{clean}}$ denote the initial faulty and healthy CER. The loop stops when no failures remain and $R_t\geq\tau$, or returns the best accepted state when its iteration budget $T$ is exhausted. Algorithm~\ref{alg:loopvsr} summarizes the procedure.

\begin{algorithm}[t]
\caption{LoopVSR repair procedure}
\label{alg:loopvsr}
\small
\begin{algorithmic}[1]
\STATE \textbf{Input:} faulty state $x_0$, repair set $\mathcal{D}_r$, clean CER, recovery target $\tau$, iteration budget $T$
\STATE Evaluate $x_0$; initialize accepted state $x_c$, feedback $o_c$, and repair history $\mathcal{H}$
\FOR{$t=1,\ldots,T$}
    \STATE Ask the agent to diagnose $o_c$ and propose patch $\Delta_t$
    \STATE Audit changed files, protected assets, and data leakage
    \IF{the candidate violates an audit constraint}
        \STATE Record the reason, roll back the patch, and \textbf{continue}
    \ENDIF
    \STATE Form candidate $x'\leftarrow x_c\oplus\Delta_t$ and run end-to-end VSR
    \STATE Collect candidate feedback $o'$: failures, telemetry, and CER
    \IF{the candidate has fewer failed samples}
        \STATE Accept it: $(x_c,o_c)\leftarrow(x',o')$
    \ELSIF{failures are equal and CER improves by more than $\epsilon$}
        \STATE Accept it: $(x_c,o_c)\leftarrow(x',o')$
    \ELSE
        \STATE Roll back to the latest accepted state
    \ENDIF
    \STATE Record the patch, feedback, and decision in $\mathcal{H}$
    \IF{no samples fail and recovery reaches $\tau$}
        \STATE \textbf{break}
    \ENDIF
\ENDFOR
\STATE Finalize $x_c$ and evaluate it on the independent hidden set
\STATE \textbf{Output:} repaired pipeline state $x^*\leftarrow x_c$
\end{algorithmic}
\end{algorithm}

Consequently, the agent need not infer every root cause from the initial observation. Each accepted state changes the executable system and may reveal evidence for the next diagnosis, while rejected candidates leave the repair trajectory unchanged. The final result is selected by audited VSR behavior and an explicit stopping rule rather than by textual patch similarity or agent self-assessment.

\section{Experiments}

\subsection{Experimental Setup}

\subsubsection{VSR System and Runtime Environment}

We use the multilingual VSR system of Ma \emph{et al.}~\cite{ma2022multilingualvsr} and evaluate its CMLR video-only model with the public CMLR weights. The model remains frozen in every experiment, with no fine-tuning, parameter update, or weight replacement. The healthy system uses input at 25 frames per second (FPS), Beam5 search, and a CTC decoding weight of 0.1. MediaPipe facial landmarks produce a $96\times96$ mouth ROI, which is center-cropped to the model input size of $88\times88$.

The code agent is Codex \texttt{gpt-5.6-sol} with \texttt{xhigh} reasoning effort. All tasks use the same system instruction and repair prompt, with no fault-specific prompting. Each task allows at most six repair iterations. Candidate patches reach real VSR inference only after file-boundary, model-path, and sample-leakage audits. We set $\epsilon=10^{-6}$ and terminate when the failure count is zero and recovery is at least 95\%.

The agent runs in an isolated local workspace, while external VSR evaluation runs on an NVIDIA Quadro RTX 8000 GPU with 48 GB memory. The execution environment uses Python, PyTorch 2.0.1, CUDA 11.7, and an identical model file. The agent cannot access the GPU server, model weights, evaluator, or hidden set; the controller alone synchronizes audited candidates to the execution environment.

\subsubsection{Data Splits}

Repair-dev contains 20 fixed videos from the CMLR training split with 210 reference characters. It is not used to train a model. Instead, it represents a low-cost engineering regression set: after every patch, the controller performs end-to-end inference on the same samples and returns failure count, CER, input statistics, and representative errors. The healthy system produces 17 character errors and 8.0952\% CER on Repair-dev.

Hidden200 contains 200 videos and 2,841 reference characters from the CMLR validation split. The healthy system obtains 11.4396\% CER. Hidden200 is never used in agent feedback, candidate selection, or stopping decisions before the patch is frozen, and is used only for final generalization evaluation.

\subsubsection{Fault Set}

We construct 13 deterministic candidate faults and freeze them before any agent execution. Inclusion in the main experiment depends only on whether a candidate causes measurable CER degradation, not on whether LoopVSR can repair it. A length-penalty candidate leaves CER unchanged, while a red--green--blue (RGB) versus blue--green--red (BGR) grayscale candidate slightly improves Repair-dev CER; both are excluded after prescreening. The fault set also covers application programming interface (API) failures in video loading and landmark detection. The final main set contains 11 tasks, summarized in Table~\ref{tab:faults}.

\begin{table*}[t]
\centering
\caption{VSR engineering faults in the experiment.}
\label{tab:faults}
\begin{tabularx}{\textwidth}{@{}l l X@{}}
\toprule
Category & Faults & Injected degradation \\
\midrule
Configuration and decoding & Input FPS, Beam size, CTC weight & Inconsistent input frame rate, reduced search width, and shifted CTC weight \\
Pixel preprocessing & Pixel scale, normalization & Incorrect pixel scaling factor and normalization statistics \\
Spatial processing & ROI size, ROI index & Reduced mouth crop and landmark indices shifted toward the eye region \\
Temporal processing & Temporal reverse, temporal stride & Reversed frame order and an additional stride-2 subsampling operation \\
Compound fault & FPS + normalization & Simultaneous configuration and preprocessing errors \\
Cascading fault & Loader + normalization & Upstream video-loading API failure masking a normalization error \\
\bottomrule
\end{tabularx}
\end{table*}

The second experiment uses three cascading stress tasks. Loader + normalization overlaps with the main set; Loader + ROI size and Detector API + temporal stride are added. An upstream API fault initially blocks all samples, hiding downstream tensor statistics and predictions. Once that fault is repaired, a normalization, spatial, or temporal fault becomes observable. The additional tasks are also frozen before agent execution.

\subsubsection{Baselines and Metrics}

We compare four system states:
\begin{itemize}
    \item Clean reference: the healthy VSR system and recovery target.
    \item Faulty system: the initial system after deterministic fault injection.
    \item Static guard: fixed rules that check input/model FPS consistency, a lower bound on beam size, the CTC-weight range, and the length-penalty range.
    \item LoopVSR: the complete diagnosis, patching, auditing, execution, and accept-or-rollback loop.
\end{itemize}
We do not add fault-specific rules to Static guard or tune the agent prompt by task. Metrics include CER, failed samples, repair success rate, recovery rate, agent iterations, patch scope, and Hidden200 inference time. A task is successfully repaired when no samples fail and $R\geq95\%$. Since insertion errors may exceed reference length, CER can exceed 100\%.

\subsection{Experiment I: Overall Repair Effectiveness}

This experiment evaluates whether LoopVSR can recover recognition quality across heterogeneous VSR engineering faults and whether it provides broader repair coverage than predefined static checks. We further examine whether accepted repairs generalize beyond the development feedback used by the controller.

\begin{figure*}[t]
\centering
\includegraphics[width=0.85\textwidth]{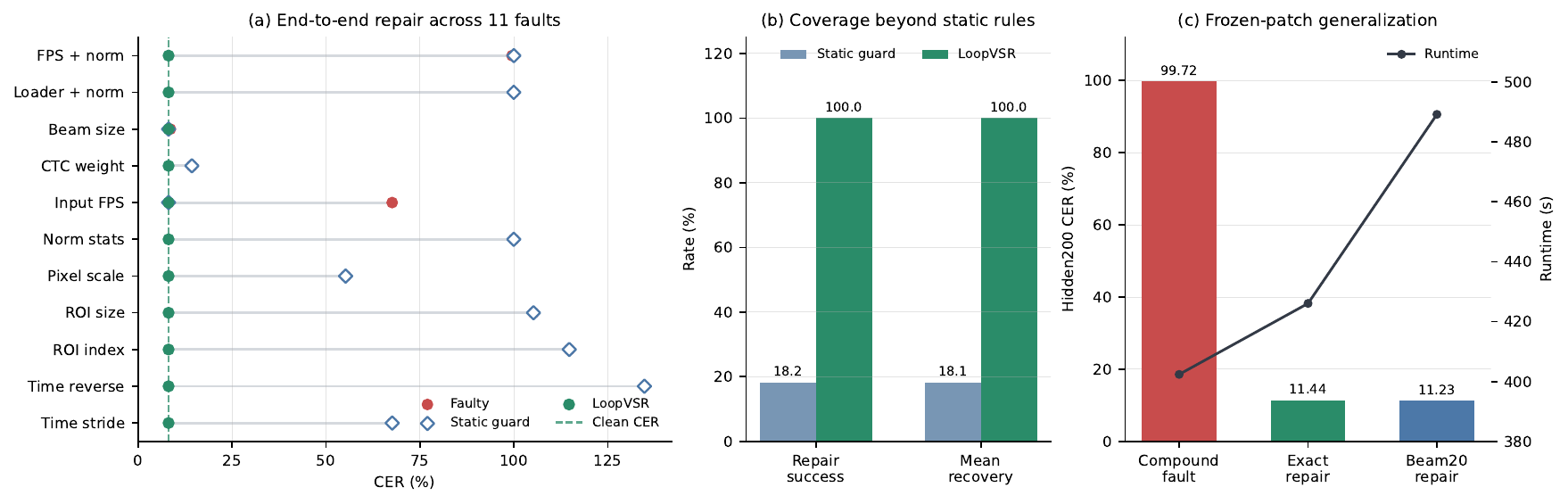}
\caption{Overall repair effectiveness. (a) CER before and after repair for 11 frozen faults; Static guard is shown as an intermediate baseline and the dashed line denotes clean CER. (b) Aggregate repair success and mean recovery. (c) Hidden200 CER and runtime for a severe compound fault, its exact repair, and the non-default Beam20 repair.}
\label{fig:repair-effectiveness}
\end{figure*}

We inject the 11 frozen faults in Table~\ref{tab:faults} and run each task on Repair-dev with the same agent prompt, six-iteration budget, patch audit, and accept-or-rollback criterion. The faulty pipeline and Static guard are evaluated against the clean CER of 8.10\%, and a repair is successful when it has zero failed samples and at least 95\% recovery. After patch selection, we freeze the accepted repair and evaluate representative runtime states on Hidden200: the severe FPS + normalization fault tests exact restoration, while the non-default Beam20 patch tests whether an alternative agent solution transfers to unseen samples. Fig.~\ref{fig:repair-effectiveness} summarizes both development-set repair and hidden-set generalization.

In Fig.~\ref{fig:repair-effectiveness}(a), the injected faults span a wide severity range, from 8.57\% CER for Beam size to 134.76\% for temporal reversal. LoopVSR restores every task to the clean 8.10\% CER, including configuration, pixel, spatial, temporal, and compound faults. In contrast, Static guard repairs only Beam size and Input FPS. Consequently, Fig.~\ref{fig:repair-effectiveness}(b) shows that Static guard achieves a 2/11 success rate (18.18\%) and 18.13\% mean recovery, whereas LoopVSR achieves 11/11 successful repairs and 100\% mean recovery. The 11 tasks require only 12 accepted iterations in total; Loader + normalization takes two iterations, and every other task takes one.

The frozen-patch results in Fig.~\ref{fig:repair-effectiveness}(c) confirm that the observed recovery is not confined to Repair-dev. On Hidden200, the compound-fault CER decreases from 99.7184\% to 11.4396\%, exactly matching the healthy runtime and yielding 100\% recovery. Beam20, the only accepted repair with a semantic runtime fingerprint different from the healthy configuration, reaches 11.2284\% CER, a 1.85\% relative reduction from Beam5. This gain increases inference time from 426.1 to 489.2 seconds (14.82\%), revealing an explicit accuracy--latency trade-off rather than development-set overfitting.

These results show that task-level execution feedback enables LoopVSR to repair faults that cannot be covered by a small collection of local validity rules. The exact compound repair transfers fully to unseen videos, while the Beam20 case demonstrates that the loop can also discover a valid non-default operating point; deployment objectives can therefore incorporate runtime cost when efficiency is required.

\subsection{Experiment II: Progressive Repair of Cascading Faults}

This experiment tests whether repeated execution is functionally necessary when one fault masks another. The key question is whether LoopVSR can use newly exposed runtime evidence to repair multiple root causes progressively rather than stopping after execution is restored.

\begin{figure}[t]
\centering
\includegraphics[width=\columnwidth]{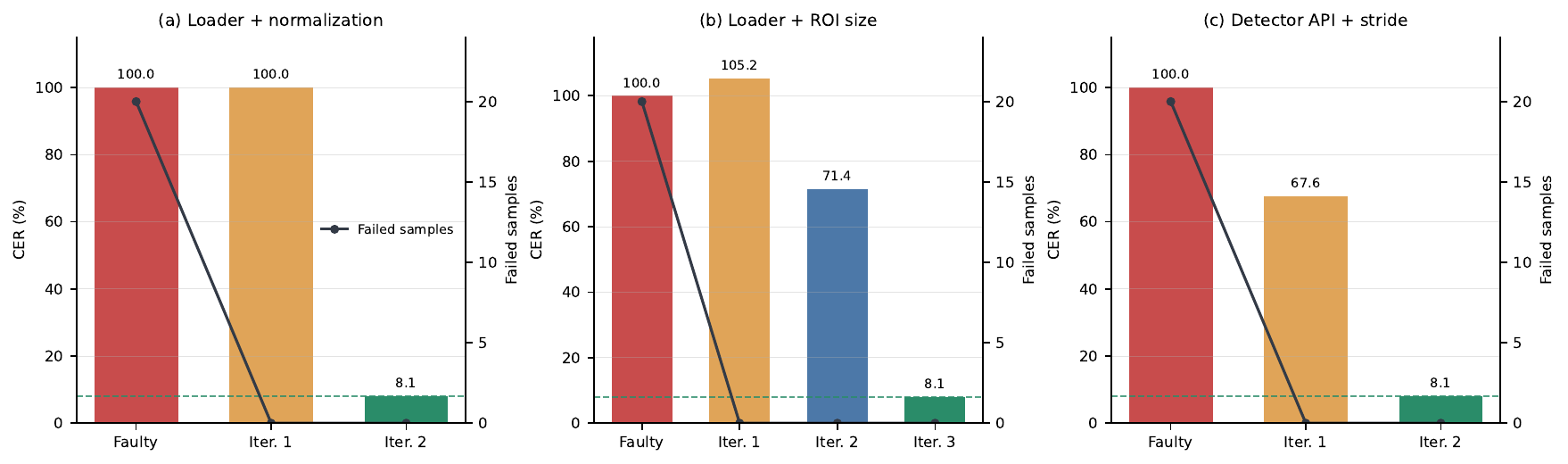}
\caption{Progressive repair of three cascading VSR faults. Bars show CER, the dark line shows failed samples, and the dashed line denotes the healthy 8.10\% CER.}
\label{fig:cascading}
\end{figure}

We construct three frozen cascading tasks on Repair-dev. In each task, an upstream loader or detector API fault initially causes all 20 samples to fail, preventing downstream tensors and predictions from being observed; the hidden downstream fault affects normalization, ROI geometry, or temporal sampling. We use the same prompt, six-iteration limit, audit rules, and acceptance criterion as in Experiment I. After every accepted patch, the controller reruns the complete VSR pipeline and returns only the observations made available by the new system state. Fig.~\ref{fig:cascading} reports the resulting failure-count and CER trajectories.

Fig.~\ref{fig:cascading} shows the same diagnostic pattern across all three tasks: the first iteration eliminates 20 execution failures but does not complete the repair. Loader + normalization remains at 100.00\% CER until the second iteration uses the newly exposed tensor range to correct normalization. Detector API + temporal stride similarly reveals shortened frame sequences only after the API call is repaired, and the second iteration restores CER from 67.62\% to 8.10\%. Loader + ROI size is the most demanding case: after execution is restored, CER first rises to 105.24\%, then falls to 71.43\% after an intermediate masking correction, and finally reaches 8.10\% when the ROI geometry is restored. Across the three tasks, LoopVSR performs seven accepted iterations and ends with zero failures, 8.10\% CER, and 100\% recovery in every case.

The cascading trajectories establish why the proposed loop is more than repeated patch generation. Each real execution changes the observable system state, exposes evidence that was unavailable in the previous iteration, and supports the next diagnosis. The final semantic fingerprints also match the healthy VSR runtime, confirming that progressive feedback recovers complete pipeline behavior rather than merely suppressing the first visible exception.

\section{Conclusion}
In this paper, we presented LoopVSR, a Loop Engineering framework for automated repair of VSR inference pipelines. LoopVSR places repository-level diagnosis and code modification within an externally controlled cycle in which audited patches are retained only after measurable end-to-end improvement. Workspace isolation and immutable model weights, evaluators, and hidden data separate the agent's repair capability from evaluation authority. This separation prevents the agent from bypassing the intended repair task and ensures that accepted patches represent genuine recovery of the VSR pipeline. Experiments covering 11 configurations, preprocessing, spatial, temporal, and compound faults restored every task to the clean 8.10\% CER, whereas Static guard succeeded on only two tasks. In three cascading cases, re-execution progressively exposed masked downstream symptoms and achieved complete recovery in seven accepted iterations. Together, these results demonstrate the effectiveness of LoopVSR in automatically diagnosing and repairing diverse VSR pipeline faults through iterative execution feedback. 

\bibliographystyle{IEEEtran}
\bibliography{ref}

@article{assael2016lipnet,
  title   = {LipNet: End-to-End Sentence-level Lipreading},
  author  = {Assael, Yannis M. and Shillingford, Brendan and Whiteson, Shimon and de Freitas, Nando},
  journal = {arXiv preprint arXiv:1611.01599},
  year    = {2016}
}

@inproceedings{shi2022avhubert,
  title     = {Learning Audio-Visual Speech Representation by Masked Multimodal Cluster Prediction},
  author    = {Shi, Bowen and Hsu, Wei-Ning and Lakhotia, Kushal and Mohamed, Abdelrahman},
  booktitle = {International Conference on Learning Representations},
  year      = {2022}
}

@article{ma2022multilingualvsr,
  title   = {Visual Speech Recognition for Multiple Languages in the Wild},
  author  = {Ma, Pingchuan and Petridis, Stavros and Pantic, Maja},
  journal = {Nature Machine Intelligence},
  volume  = {4},
  pages   = {930--939},
  year    = {2022},
  doi     = {10.1038/s42256-022-00550-z}
}

@inproceedings{ma2023autoavsr,
  title     = {Auto-AVSR: Audio-Visual Speech Recognition with Automatic Labels},
  author    = {Ma, Pingchuan and Haliassos, Alexandros and Fernandez-Lopez, Adriana and Chen, Honglie and Petridis, Stavros and Pantic, Maja},
  booktitle = {IEEE International Conference on Acoustics, Speech and Signal Processing},
  pages     = {1--5},
  year      = {2023},
  doi       = {10.1109/ICASSP49357.2023.10096889}
}

@article{legoues2012genprog,
  title   = {GenProg: A Generic Method for Automatic Software Repair},
  author  = {Le Goues, Claire and Nguyen, ThanhVu and Forrest, Stephanie and Weimer, Westley},
  journal = {IEEE Transactions on Software Engineering},
  volume  = {38},
  number  = {1},
  pages   = {54--72},
  year    = {2012},
  doi     = {10.1109/TSE.2011.104}
}

@inproceedings{nguyen2013semfix,
  title     = {SemFix: Program Repair via Semantic Analysis},
  author    = {Nguyen, Hoang Duong Thien and Qi, Dawei and Roychoudhury, Abhik and Chandra, Satish},
  booktitle = {International Conference on Software Engineering},
  pages     = {772--781},
  year      = {2013},
  doi       = {10.1109/ICSE.2013.6606623}
}

@inproceedings{wardat2022deepdiagnosis,
  title     = {DeepDiagnosis: Automatically Diagnosing Faults and Recommending Actionable Fixes in Deep Learning Programs},
  author    = {Wardat, Mohammad and Cruz, Breno Dantas and Le, Wei and Rajan, Hridesh},
  booktitle = {IEEE/ACM International Conference on Software Engineering},
  pages     = {561--572},
  year      = {2022},
  doi       = {10.1145/3510003.3510071}
}

@inproceedings{yao2023react,
  title     = {ReAct: Synergizing Reasoning and Acting in Language Models},
  author    = {Yao, Shunyu and Zhao, Jeffrey and Yu, Dian and Du, Nan and Shafran, Izhak and Narasimhan, Karthik and Cao, Yuan},
  booktitle = {International Conference on Learning Representations},
  year      = {2023}
}

@inproceedings{shinn2023reflexion,
  title     = {Reflexion: Language Agents with Verbal Reinforcement Learning},
  author    = {Shinn, Noah and Cassano, Federico and Gopinath, Ashwin and Narasimhan, Karthik and Yao, Shunyu},
  booktitle = {Advances in Neural Information Processing Systems},
  volume    = {36},
  year      = {2023}
}

@inproceedings{jimenez2024swebench,
  title     = {SWE-bench: Can Language Models Resolve Real-world GitHub Issues?},
  author    = {Jimenez, Carlos E. and Yang, John and Wettig, Alexander and Yao, Shunyu and Pei, Kexin and Press, Ofir and Narasimhan, Karthik},
  booktitle = {International Conference on Learning Representations},
  year      = {2024}
}

@inproceedings{yang2024sweagent,
  title     = {SWE-agent: Agent-Computer Interfaces Enable Automated Software Engineering},
  author    = {Yang, John and Jimenez, Carlos E. and Wettig, Alexander and Lieret, Kilian and Yao, Shunyu and Narasimhan, Karthik and Press, Ofir},
  booktitle = {Advances in Neural Information Processing Systems},
  volume    = {37},
  year      = {2024},
  doi       = {10.52202/079017-1601}
}

@inproceedings{zhang2024autocoderover,
  title     = {AutoCodeRover: Autonomous Program Improvement},
  author    = {Zhang, Yuntong and Ruan, Haifeng and Fan, Zhiyu and Roychoudhury, Abhik},
  booktitle = {ACM SIGSOFT International Symposium on Software Testing and Analysis},
  year      = {2024},
  doi       = {10.1145/3650212.3680384}
}

@inproceedings{bouzenia2025repairagent,
  title     = {RepairAgent: An Autonomous, LLM-Based Agent for Program Repair},
  author    = {Bouzenia, Islem and Devanbu, Premkumar and Pradel, Michael},
  booktitle = {IEEE/ACM International Conference on Software Engineering},
  pages     = {2188--2200},
  year      = {2025},
  doi       = {10.1109/ICSE55347.2025.00157}
}

@misc{openai2025codex,
  title        = {Introducing Codex},
  author       = {{OpenAI}},
  howpublished = {OpenAI},
  year         = {2025},
  month        = may,
  url          = {https://openai.com/index/introducing-codex/}
}

@misc{anthropic2025claudecode,
  title        = {Claude 3.7 Sonnet and Claude Code},
  author       = {{Anthropic}},
  howpublished = {Anthropic},
  year         = {2025},
  month        = feb,
  url          = {https://www.anthropic.com/news/claude-3-7-sonnet}
}

@misc{anthropic2026loops,
  title        = {Loop Engineering: Getting Started with Loops},
  author       = {de Oliveira, Delba and Segner, Michael},
  howpublished = {Anthropic Claude Blog},
  year         = {2026},
  month        = jun,
  url          = {https://claude.com/blog/getting-started-with-loops}
}

\end{document}